\documentclass[aps,prd,twocolumn,amsmath,amssymb]{revtex4-2}
\usepackage{epsfig,xcolor,amsmath,tikz}
\usepackage{graphicx}
\usepackage{epsfig}
\usepackage{color}
\usepackage{physics}
 \begin{document}
\title{Generalized Unimodular Gravity and Cosmological Perturbations}
\author{J\'ulio C. Fabris}
\email[]{julio.fabris@cosmo-ufes.org}
\affiliation{N\'ucleo Cosmo-ufes \& Departamento de F\'isica, Universidade Federal do Esp\'irito Santo,
Avenida Fernando Ferrari, 514, Goiabeiras, 29060-900, Vit\'oria, Espir\'ito Santo, Brazil }
\author{Alexander Yu. Kamenshchik}
\email[]{kamenshchik@bo.infn.it}
\affiliation{Department of Physics and Astronomy ``A. Righi'', University of Bologna, via Irnerio 46,  40126 Bologna, Italy\\
INFN, section of Bologna, viale Berti Pichat 6/2, 40127 Bologna, Italy}
\begin{abstract}
The generalized unimodular theory is revisited and its consequence for cosmology is discussed. The usual matter components of the universe are obtained in a pure geometric way. This result gives a new perspective to the studies of the dark sector of the universe.
A background and perturbative analysis are carried out, recovering the corresponding results obtained through the general relativity theory but with a different interpretation.
\end{abstract}
\maketitle

\section{Introduction}

The idea of the unimodular gravity was put forward already by Einstein as early as in 1919 \cite{Einstein}. Later it was further developed in papers \cite{Unruh} and \cite{Henneaux} and in many others. It can be formulated in many different ways. One can, for example, require that the theory were invariant with respect not to all the diffeomorphisms, but only with respect to the so called volume-preserving or area-preserving or divergenceless diffeomorphisms, i.e. by the diffeomorphisms generated by the divergenceless transverse vector fields. In such a model the super-Hamiltonian constraint is also not satisfied and in the analogous first-order equation arising as a first integral of other (second-order) Einstein equations we find an integration constant playing the role of the cosmological constant. The unimodular gravity  attracts a growing attention of researchers during last years.    

In the paper \cite{we}, the theory of the generalized unimodular gravity was developed. In this approach the unimodular gravity is represented as follows: the lapse function is treated non as a free Lagrange multiplier but as a function of the determinant of the spatial part of the metric. Namely, we can fix the condition 
\begin{equation}
N = \frac{1}{\gamma}.
\label{unimod0}
\end{equation}
This condition is equivalent to the condition
\begin{equation}
g = -1,
\label{unimod1} 
\end{equation}
where 
$N$ is the lapse function, $\gamma$ is the determinant of the three-dimensional spatial metric and $g$ is the determinant of the 
spacetime metric. The fact that the determinant $g$ is constant is equivalent to the requirement that the theory is invariant with respect to the group of area-preserving diffeomorphisms, and hence, we reproduce the unimodular gravity. One can take the expression \eqref{unimod0} and to substitute it into the spatial Einstein equations.
The resulting system of equations has the first integral which coincides with the super-Hamiltonian constraint in which the integration constant playing the role of the cosmological term is included. Then, a natural generalization arises: instead of the the \eqref{unimod0}, one defines the lapse function as another function of the determinant 
of the spatial metric, then discards the super-Hamiltonian and substitute the function
\begin{equation}
N = N(\gamma)
\label{unimod2}
\end{equation}
into the second-order Einstein equations, one obtains another first integral, including different kinds of matter, whose form depends on the form of the function $N(\gamma)$.
This generalized unimodular gravity was formulated in \cite{we}. Different applications of the generalized unimodular gravity were developed in papers \cite{Barv, Barv1, Kam-Tron-Ven,Barv-Kol-Vik,Barv-Nest,Nest-Lyam,Nest}. Let us note that if one chooses the function 
\begin{equation}
N = 1,
\label{Burlan}
\end{equation}
then one obtains the first integral of the Einstein equations, where the additional dust-like matter is present. In another context this case was considered in the book \cite{Burlan} and the series of papers cited therein.

The relaxation of a part of constraints which generates additional ``matter'' degrees of freedom was used not only in the gravity theory. One of the examples is the Dirac's non-linear electrodynamics \cite{Dirac}. As is well known the only constraint in the electrodynamics is the Gauss law 
\begin{equation}
{\rm div} \vec{E} = \rho,
\label{Dirac}
\end{equation}  
where $\vec{E}$ is an electric field and $\rho$ is the electric charge density.   The electrodynamics without this constraint was considered by Dirac in 1951 \cite{Dirac}. It was shown  that discarding this constraint, one acquires an additional degree of freedom which can be identified with the electron.  The ideas of Dirac's new non-linear electrodynamics were further developed by many authors, (see e.g. \cite{post,post1,post2,post3}).

A somewhat  different approach to the treatment of the systems with constraints called ``unfree gauge symmetry'' was developed in the series of papers \cite{Semen,Semen1,Semen2,Semen3}. There the systems with first class constraints are treated as the systems with gauge symmetries. Then, if one instead of arbitrary spacetime gauge functions considers the functions which satisfy additional conditions, namely, they satisfy some partial differential equations,  the new theories arise. The simplest example of such a theory is again the unimodular gravity, where the corresponding equations are simply the conditions of the transverseness of diffeomorphisms:
 \begin{equation}
 \nabla_i\xi^i = 0. 
 \label{trans}
 \end{equation}
 
 The simplest constrained system is a relativistic particle. In the series of  papers by Stueckelberg \cite{Stueck,Stueck1,Stueck2} the quantization of such a particle with the mass which is not fixed was considered. In this case the mass plays a role similar to that played by the cosmological constant in the unimodular gravity theories. Thus, relaxing the constraint (on shell condition) one acquires an additional degree of freedom. 
 
 We would like to mention also the so called TDiff theories \cite{Tdiff}.  In this approach the matter part of the action is invariant only with respect to the transverse diffeomorphisms. Thus, it looks like some weaker version of the unimodular gravity. Similar models were studied also in paper \cite{Tdiff1}, where the details of the breaking of the diffeomorphism invariance were considered.  Rather a broad analysis of the gravity theories with different sets of constraints or different structures of gauge algebras was presented in paper \cite{Percacci}. In paper \cite{Magueijo} it was shown how the degradation of the  invariance 
 with respect to the spacetime diffeomorphisms to the subgroup of the spatial diffeomorphisms implies the appearance of new degrees of freedom. 
 
 Generally, one can say that in a theory with gauge symmetry or in the theory with constraints \cite{constrained}, the cancellation of some constraints implies the appearance of new physical degrees of freedom. Attempts to analyse this phenomenon from a general point of view were undertaken in papers \cite{we-all,Golovnev}. 

In the present paper we analyze the applications of the generalized unimodular gravity to cosmology. We use the most direct approach, treating the lapse function not a Lagrange multiplier, but as a function of the determinant of the spatial metric. 
As a result we have 9 Einstein equations instead of 10, and these 9 equations are modified by some additional terms. 
Then combining these 9 modified Einstein equations with the Bianchi identities, we obtain 10th equation, which represent a $00$ component of the Einstein equations which contains an additional term mimicking some kind of matter. We show how it works for a simple Friedmann universe, then we derive some general formulas and finally, we study the cosmological perturbations in the generalized unimodular cosmology. The main goal of our investigations is to understand if there are some (in principle) observable effects which distinguish the generalized unimodular gravity from General Relativity. The structure of the paper is the following one. In next section the consequences of imposing the general condition (\ref{unimod2}) for the unperturbed cosmological configuration are determined. In section III, the perturbations on the resulting cosmological scenario are studied. In section IV we draw our conclusions.

\section{Einstein equations and Bianchi identities}

In this section we shall present the general formalism of the generalized unimodular gravity in terms of the of the modification of the Einstein equations. The use of the Bianchi identities will be rather useful, because they have purely geometric character and are always valid, independently on the gravity theory.

Let us consider the Hilbert-Einstein action in the empty spacetime. We shall look for the solutions 
which can be represented in the comoving frame coordinates, where the space-time components of the metric are equal to zero. Thus, the metric has the following form:
\begin{equation}
ds^2 = N^2dt^2-\gamma_{\alpha\beta}dx^{\alpha}dx^{\beta}.
\label{metric}
\end{equation}
We suppose that the lapse function $N(t)$ is not a independent field variable but a fixed function of the determinant of the spatial metric:
\begin{equation}
N = N(\gamma).
\label{N}
\end{equation}
That means that we do not have the $00$ component of the Einstein equation, or, in other words, the super-Hamiltonian constrain is absent and we have 9 equations instead of 10 ones. However, the $\alpha-\beta$ component of the Einstein equations is modified because we should add to the term $\frac{\delta S}{\delta g_{ij}}$ the term $\frac{\delta S}{\delta g_{00}}\frac{\delta g_{00}}{\delta g_{\alpha\beta}}$. As a result we obtain the following equation:
\begin{equation}
R_{\alpha}^{\beta}-\frac12\delta_{\alpha}^{\beta}R+\frac{2N'\gamma}{N}\delta_{\alpha}^{\beta}\left(R_0^0-\frac12 R\right)=0.
\label{Einstein}
\end{equation} 
Here, the ``prime'' denotes the differentiation with respect to $\gamma$ and we have used the fact that for the metric \eqref{metric} $g_{00} = N^2$ and $g_{\alpha\beta} = -\gamma_{\alpha\beta}$. 
This equation can be treated as a spatial-spatial Einstein equation with the nonvanishing right-hand side proportional to 
the expression 
\begin{equation}
R_0^0-\frac12 R,
\label{00}
\end{equation}
which is now non obliged to be equal to zero and can play a role of the effective energy density. 
It will be convenient start our considerations with analysis of the simplest case - the cosmology of the flat Friedmann universe. 

\subsection{Flat Friedmann universe}

Let us consider the flat Friedmann metric
\begin{equation}
ds^2 = N^2(t)dt^2 - a^2(t)(dx^2+dy^2+dz^2).
\label{Fried}
\end{equation}
For this metric, the Christoffel symbols are  
\begin{equation}
\Gamma_{00}^{0} = \frac{\dot{N}}{N}, \Gamma_{\alpha\beta}^0=\frac{\dot{a}a}{N^2}\delta_{\alpha\beta}, \Gamma_{\alpha 0}^{\beta}=\delta_{\alpha}^{\beta}\frac{\dot{a}}{a}.
\label{Chris}
\end{equation}
The ``dot'' means the differentiation with respect to the time parameter $t$.
The nonvanishing components of the Ricci tensor are 
\begin{equation}
R_{\alpha}^{\beta}=\delta_{\alpha}^{\beta}\left(-\frac{\ddot{a}}{aN^2}-2\frac{\dot{a}^2}{a^2N^2}+\frac{\dot{N}\dot{a}}{N^3a}\right),
\label{Ricci}
\end{equation}
\begin{equation}
\label{Ricci1}
R_{0}^{0}=-3\frac{\ddot{a}}{aN^2}+3\frac{\dot{N}\dot{a}}{N^3a}.
\end{equation}
The Ricci scalar is 
\begin{equation}
R = -6\frac{\ddot{a}}{aN^2}+6\frac{\dot{N}\dot{a}}{N^3a}-6\frac{\dot{a}^2}{N^2a^2}.
\label{scal}
\end{equation}
Substituting the expressions \eqref{Ricci}, \eqref{Ricci1} and \eqref{scal} into Eq. \eqref{Einstein}, we obtain
the following equation:
\begin{equation}
2\frac{\ddot{a}}{aN^2}+\frac{\dot{a}^2}{a^2N^2}-2\frac{\dot{N}\dot{a}}{N^3a}+6\frac{N'\gamma\dot{a}^2}{a^2N^3}=0.
\label{Fried1}
\end{equation}
Noting that 
\begin{equation}
\dot{N} = N'\dot{\gamma}=6N'\dot{a}a^5,
\label{N1}
\end{equation}
we reduce Eq. \eqref{Fried1} as follows:
\begin{equation}
2\frac{\ddot{a}}{aN^2}+\frac{\dot{a}^2}{a^2N^2}-\frac{\dot{N}\dot{a}}{N^3a}=0.
\label{Fried2}
\end{equation}
Now, we introduce the Hubble parameter with respect to the cosmic time:
\begin{equation}
H \equiv \frac{a_{\tau}}{a}=\frac{\dot{a}}{aN},
\label{Hubble}
\end{equation}
where 
\begin{equation}
d\tau = Ndt.
\label{cosmic}
\end{equation}
We shall try to find the first integral the second order equation \eqref{Fried2}. 
Let us write down a chain of identities:
\begin{equation}
\dot{a} = aNH=a_{\tau}N,
\label{iden}
\end{equation}
\begin{equation}
\ddot{a} = \dot{a}NH+a\dot{N}H+aN\dot{H}=a_{\tau}N^2H+aN_{\tau}NH+aN^2H_{\tau}.
\label{iden1}
\end{equation}
Substituting the formulas \eqref{iden} and \eqref{iden1} into Eq. \eqref{Fried2}, we obtain
\begin{equation}
2H_{\tau}+3H^3+\frac{N_{\tau}H}{N} = 0.
\label{Fried3}
\end{equation}
Dividing Eq. \eqref{Fried3} by $H$, we obtain
\begin{equation}
\frac{d\ln H^2}{d\tau}+3H+\frac{d\ln N}{d\tau}=0.
\label{Fried4}
\end{equation}
The first integral of this equation is 
\begin{equation}
H^2a^3N = C = constant.
\label{Fried5}
\end{equation}
We can rewrite Eq. \eqref{Fried5} in the standard Friedmann form
\begin{equation}
H^2 = \frac{C}{Na^3}.
\label{Fried6}
\end{equation}
Thus, we see that due to the fact that $N$ is not an independent metric variable but a given function of $\gamma$ (i.e. of $a$ in this simplest case), the constant $C$ is not obliged to be equal to zero and, hence, an effective matter with the energy density 
\begin{equation}
\varepsilon = \frac{C}{Na^3},
\label{energy}
\end{equation}
arises. If 
\begin{equation}
N = 1
\label{dust}
\end{equation}
or to some other constant, this effective matter behaves as dust. If 
\begin{equation}
N = \frac{1}{\sqrt{\gamma}} = \frac{1}{a^3},
\label{unimod}
\end{equation}
one obtains an effective cosmological constant, i.e. one reproduces the unimodular gravity theory. 
If 
\begin{equation}
N = \gamma^{\frac{w}{2}},
\label{barotrop}
\end{equation}
where $w$ is a constant, 
we obtain the effective perfect fluid with the equation of state 
\begin{equation}
p = w\varepsilon,
\label{pressure}
\end{equation}
where $p$ is the pressure.
Using Eq. \eqref{energy}, one can easily obtain the forms of the lapse function reproducing the behavior of more complicated Friedmann models. For example, the lapse function
\begin{equation}
N = \frac{1}{1+C_1 a^3} = \frac{1}{1+C_1\sqrt{\gamma}},
\label{mixture}
\end{equation}
describes a Friedmann universe filled with the mixture of dust and cosmological constant. 

Now, the next question arises: what can we do with Eq. \eqref{Einstein} in a more general case. We shall study the general metric but keeping the equation $g_{0\alpha} = 0$ intact.

\subsection{Remaining in the comoving frame}

Let us consider the Bianchi identity 
\begin{equation}
\nabla_{i}\left(R_0^i-\frac12\delta_0^iR\right)=0.
\label{Bianchi}
\end{equation}
In our case it is reduced to 
\begin{equation}
\partial_{0}\left(R_0^0-\frac12R\right)=\Gamma_{0\beta}^{\alpha}R_{\alpha}^{\beta}-\Gamma_{0\alpha}^{\alpha}R_0^0,
\label{Bianchi1}
\end{equation}
 where the Christoffel symbols are
 \begin{equation}
 \Gamma_{0\beta}^{\alpha} = \frac12\gamma^{\alpha\delta}\dot{\gamma}_{\delta\beta},\ \Gamma_{0\alpha}^{\alpha}= 
 \frac12\gamma^{\alpha\delta}\dot{\gamma}_{\delta\alpha}.
 \label{Chris1}
 \end{equation}
 It follows from Eq. \eqref{Einstein} that 
 \begin{equation}
 R_{\alpha}^{\beta} =\frac12\delta_{\alpha}^{\beta}R-\frac{2N'\gamma}{N}\delta_{\alpha}^{\beta}\left(R_0^0-\frac12 R\right).
 \label{Einstein1}
 \end{equation}
Substituting the expressions \eqref{Chris1} and \eqref{Einstein1} into Eq. \eqref{Bianchi1}, we obtain
\begin{eqnarray}
&&\partial_{0}\left(R_0^0-\frac12R\right)\nonumber \\
&&=\gamma^{\alpha\delta}\dot{\gamma}_{\alpha\delta}\left(\frac14R-\frac12R_0^0-
\frac{N'\gamma}{N}\left(R_0^0-\frac12R\right)\right).
\label{Bianchi2}
\end{eqnarray}
Introducing the notation
\begin{equation}
\varepsilon = R_0^0-\frac12R
\label{energy1}
\end{equation}
and using the identity
\begin{equation}
\delta \gamma = \gamma\gamma^{\alpha\beta}\delta\gamma_{\alpha\beta},
\label{iden2}
\end{equation}
we can rewrite Eq. \eqref{Bianchi2} in the following simple form 
\begin{equation}
\dot{\varepsilon}=-\frac12\varepsilon\frac{\dot{\gamma}}{\gamma}-\dot{\gamma}\varepsilon\frac{N'}{N}.
\label{Bianchi3}
\end{equation}
Integrating this equation, we find that
\begin{equation}
\varepsilon = \frac{C(x^{\alpha})}{\sqrt{\gamma}N}.
\label{energy2}
\end{equation}
Here $C$ is a function of the spatial coordinates. The right-hand side of Eq. \eqref{energy2} represents a generalization of the formula \eqref{energy} and can be interpreted as an effective energy density. Analogously, the equation of state parameter is
\begin{equation}
w = 2\frac{d\ln N}{d\ln \gamma}.
\label{pressure1}
\end{equation}
However, we should understand if it is necessary to impose some additional conditions on the function $C$, and perhaps on the very form of the lapse function $N$ treated as a function of $\gamma$. 

Let us consider  another Bianchi identity:
\begin{equation}
\nabla_iT_{\alpha}^i=0. 
\label{Bianchi5}
\end{equation}
It can be rewritten as
\begin{equation}
p_{,\alpha} + 
\Gamma_{0\alpha}^0(p+\varepsilon)=0,
\label{Bianchi6}
\end{equation}
where 
\begin{equation}
\Gamma_{0\alpha}^0 = \frac{N_{,\alpha}}{N}.
\label{Chris2}
\end{equation}
We begin with some simple cases. If $N=1$, the pressure is equal zero and Eq. \eqref{Bianchi6} does not impose further restrictions on the form of the function $C$. Another simple case is $w=-1$. In this case the pressure should not depend on the spatial coordinates and that means that the same is true for the energy density and one finally comes back to the standard unimodular gravity. 

Now, let us consider the case, when the function $C$ is a constant. Then, Eq. \eqref{Bianchi6} is reduced to the following relation:
\begin{equation}
\gamma_{,\alpha}\left[\left(\frac{2N'\sqrt{\gamma}}{N^2}\right)'+\frac{N'}{N}\left(\frac{2N'\sqrt{\gamma}}{N^2}+\frac{1}{N\sqrt{\gamma}}\right)\right]=0.
\label{Bianchi7}
\end{equation}
If $\gamma_{,\alpha} = 0$ we come back to the Friedmann universe. If $\gamma_{,\alpha} \neq 0$, when the expression in the squared brackets of the expression \eqref{Bianchi7} should be equal to zero and it implies that 
\begin{equation}
\frac{N'\gamma}{N} = n = const,
\label{Bianchi8}
\end{equation}
i.e. we restrict the possible form of the function $N$, which should satisfy the law
\begin{equation}
N = N_0 \gamma^n.
\label{power}
\end{equation}

If we instead permit the non-trivial dependence of $C$ on the spatial coordinates, then we obtain the following first integral from Eq. \eqref{Bianchi6}:
\begin{equation}
\frac{CN'\gamma}{N} = C_0 = const
\label{Bianchi9}
\end{equation}
and this again send us to the situation where $N$ is described by the function \eqref{power} and $C$ is a constant.
The dependence of $N$ on $\gamma$ described by Eq. \eqref{power} means, that in the corresponding versions of the generalized unimodular gravity we can mimic only the appearance of the barotropic perfect fluids with the equations of state
$p = w\varepsilon$, where $w$ is constant. This looks rather restrictive because we cannot consider the functions similar to \eqref{mixture}, presented at the end of the preceding subsection. We shall try to find a way out from this situation.  

Now, we would like to consider a more general situation, when the shift functions are not put equal to zero, and the velocity of an effective fluid is not vanishing. Can we do it, treating the shift function as an independent field variable? We shall try to answer this question in the next subsection.

\subsection{Effective fluid in a non-comoving frame}

Calculating the variation of  the Hilbert - Einstein action with respect to the metric components $g_{0\alpha}$, or, in other words, with respect to the shift function components $N_{\alpha}$, we obtain the equation 
\begin{equation}
R^{0\alpha} - \frac12g^{0\alpha}R = 0. 
\label{velocity-a}
\end{equation}
This implies that the spatial  velocity of the effective fluid should be equal to zero. To be able to produce the effective fluid with non-vanishing velocity, we can try to treat the components of the shift function $N_{\alpha}$ as some functions of the spatial metric. Such an approach represents in a way some further modification of the generalized unimodular gravity.   
In this case Eq. \eqref{velocity-a} does not follow from the variation of the action, while Eq. \eqref{Einstein} undergoes 
an additional  modification. Namely, the term proportional to the expression to $R^{0\alpha} - \frac12g^{0\alpha}R$ should be added. Undergoing the variation of the action with respect to the $g_{\alpha\beta}=-\gamma_{\alpha\beta}$, we obtain:
\begin{widetext}
\begin{equation}
-\left(R^{\alpha\beta}-\frac12g^{\alpha\beta}R\right) + \left(R^{00}-\frac12g^{00}R\right)\frac{\delta g_{00}}{\delta \gamma_{\alpha\beta}}+2\left(R^{0\gamma}-\frac12g^{0\gamma}R\right)\frac{\delta N_{\gamma}}{\delta \gamma_{\alpha\beta}}= 0. 
\label{Einstein-new}
\end{equation}

It is more convenient to rewrite this expression in terms of the Einstein tensor $G^{ij}=R^{ij}-\frac12g^{ij}R$:
\begin{eqnarray}
G^{\alpha\beta}=G^{00}\left(2N'N\gamma\gamma^{\alpha\beta}+N^{\alpha}N^{\beta}-2N^{\gamma}\frac{d N_{\gamma}}{d\gamma_{\alpha\beta}}\right)
+2G^{0\gamma}\frac{d N_{\gamma}}{d\gamma_{\alpha\beta}}.
\label{Einstein-new1}
\end{eqnarray}
We shall write down the Bianchi identity in the form
\begin{equation}
\label{Bianchi-new}
G^{ij}_{;i}=0.
\end{equation}
In the case $j=0$ it looks like

\begin{equation}
G^{00}_{,0}+G^{0\alpha}_{,\alpha}+(2\Gamma_{00}^{0}+\Gamma_{\alpha 0}^{\alpha})G^{00}+(3\Gamma_{\alpha 0}^{\alpha}+\Gamma_{\beta\alpha}^{\beta})G^{0\alpha}+\Gamma_{\alpha\beta}^0G^{\alpha\beta}=0. 
\label{Bianchi-new1}
\end{equation}
For the spatial index $j = \alpha$, we have
\begin{equation}
G^{0\alpha}_{,0}+G^{\alpha\beta}_{,\beta}+(\Gamma_{00}^0+\Gamma_{0\beta}^{\beta})G^{0\alpha}+
2\Gamma_{0\beta}^{\alpha}G^{0\beta}+\Gamma_{00}^{\alpha}G^{00}+\Gamma_{\beta\gamma}^{\beta}G^{\gamma\alpha}+\Gamma_{\beta\gamma}^{\alpha}G^{\beta\gamma}=0.
\label{Bianchi-new2}
\end{equation}
\end{widetext}
To check the correctness of Eqs. \eqref{Einstein-new}, \eqref{Bianchi-new1} and \eqref{Bianchi-new2} let us consider the preceding case, when $N_{\alpha}=0$ and $G^{0\alpha}=0$.
Then Eq. \eqref{Einstein-new} becomes 
\begin{equation}
G^{\alpha\beta}=2N'N\gamma\gamma^{\alpha\beta}G^{00},
\label{Einstein-new1}
\end{equation}  
while Eq. \eqref{Bianchi-new1} is 
\begin{equation}
G^{00}_{,0}+(2\Gamma_{00}^0+\Gamma_{0\alpha}^{\alpha})G^{00}+2\Gamma_{\alpha\beta}^0N'N\gamma\gamma^{\alpha\beta}G^{\alpha\beta}=0.
\label{Bianchi-new3}
\end{equation}
Using the expressions for the Christoffel symbols \eqref{Chris1} and substituting the expression \eqref{Einstein-new1} into Eq. 
\eqref{Bianchi-new3}, we obtain the following equation:
\begin{equation}
G^{00}_{,0}+3\frac{\dot{N}}{N}G^{00}+\frac12\frac{\dot{\gamma}}{\gamma}G^{00}=0.
\label{Bianchi-new4}
\end{equation}
Integrating this equation, we find
\begin{equation}
G^{00}=\frac{C(x^{\alpha})}{N^3\sqrt{\gamma}},
\label{G-00}
\end{equation}
and, hence, 
\begin{equation}
G_{0}^0 = \frac{C(x^{\alpha})}{N\sqrt{\gamma}},
\label{G-001}
\end{equation}
which coincides with the expression \eqref{energy2}, as it should be.

Now, we would like to find a solution of the system of equations \eqref{Einstein-new}, \eqref{Bianchi-new1} and \eqref{Bianchi-new2}
such that the corresponding functions $G^{ij}$ can be represented in the form
\begin{equation}
G^{ij} = (\varepsilon+p)u^{i}u^j-g^{ij}p.
\label{standard}
\end{equation}
A reasonable choice for the velocities is 
\begin{equation}
u^0=\frac{1}{N},\ u^{\alpha}=\frac{N^{\alpha}}{N}.
\label{velocity}
\end{equation}
Substituting the expressions \eqref{velocity} into Eq. \eqref{standard} and substituting the obtained formula into Eq. \eqref{Einstein-new}, we obtain
\begin{equation}
\left(\frac{2N'\gamma\gamma^{\alpha\beta}}{N}\right)\varepsilon=\left(\gamma^{\alpha\beta}-\frac{2N^{\gamma}}{N^2}\frac{dN_{\gamma}}{d\gamma_{\alpha\beta}}\right)p.
\label{Einstein-new2}
\end{equation}
To make this relation reasonable, we should require that the expression $\frac{2N^{\gamma}}{N^2}\frac{dN_{\gamma}}{d\gamma_{\alpha\beta}}$ be proportional to $\gamma^{\alpha\beta}$.
We suppose that 
\begin{equation}
\vec{N}^2 = N^{\gamma}N_{\gamma}  = \gamma^{\alpha\beta}N_{\alpha}N_{\beta} 
\label{ansatz}
\end{equation}
is a function of the determinant $\gamma$. Then 
\begin{equation}
\delta \vec{N}^2 = -N^{\alpha}N^{\beta}\delta\gamma_{\alpha\beta}+2N^{\gamma}\frac{dN_{\gamma}}{d\gamma_{\alpha\beta}}\delta\gamma_{\alpha\beta}=
(\vec{N}^2)'\gamma\gamma^{\alpha\beta}\delta\gamma_{\alpha\beta}.
\label{ansatz1}
\end{equation}
Then,
\begin{equation}
2N^{\gamma}\frac{dN_{\gamma}}{d\gamma_{\alpha\beta}} = (\vec{N}^2)'\gamma\gamma^{\alpha\beta}+N^{\alpha}N^{\beta}.
\label{ansatz2}
\end{equation}
We suppose also that 
\begin{equation}
N^{\alpha}N^{\beta} = \frac13\vec{N}^2\gamma^{\alpha\beta}.
\label{ansatz3}
\end{equation}
Now, Eq. \eqref{Einstein-new2} gives
\begin{equation}
p =\left(\frac{\frac{2N'\gamma}{N}}{1-\frac{(\vec{N}^2)'\gamma}{N^2}-\frac{\vec{N}^2}{3N^2}}\right)\varepsilon.
\label{w-new}
\end{equation}
We shall substitute the obtained expressions into Eq. \eqref{Bianchi-new1} to try to find the form of the function $\varepsilon$.
Let us introduce a new expression for $w$:
\begin{equation}
w = \frac{\frac{2N'\gamma}{N}}{1-\frac{(\vec{N}^2)'\gamma}{N^2}-\frac{\vec{N}^2}{3N^2}}.
\label{w-new}
\end{equation}
Now we have 
\begin{equation}
G^{00}=\frac{\varepsilon}{N^2},
\label{G-00-new}
\end{equation}
\begin{equation}
G^{0\alpha}=\frac{\varepsilon(1+w)N^{\alpha}}{N^2},
\label{G-0a}
\end{equation}
\begin{equation}
G^{\alpha\beta}=\frac{\varepsilon(1+3w)\gamma^{\alpha\beta}}{3}.
\label{G-ab}
\end{equation}

The statement \eqref{ansatz3} cannot be correct. The only reasonable suggestion is that 
\begin{equation}
N_{\alpha} = const.
\label{hypothesis}
\end{equation}
However, this option is not good as well.

Thus, we have seen that if we would like to consider a geometry which is more complicated than the homogeneous and isotropic, the only consistent choice for the lapse function $N$ is that given by Eq. \eqref{power}, which implies the appearance of effective barotropic fluid with the pressure which is proportional to the energy density, i.e. its equation of state parameter $w$ is constant. 
In what follows we shall study only such models. 

\section{Cosmological perturbations in the gauge-invariant formalism}

In this section we shall study the cosmological perturbations on the background of the flat Friedmann geometry. We shall analyse if there is a difference between their behavior in the generalized unimodular gravity and in General Relativity. 
We shall use the gauge-invariant formalism, suggested in \cite{Bardeen} (see also \cite{Mukhanov-PR,Mukhanov-book}).
Let us write the metric in the following form, taking into account only the scalar perturbations:
\begin{equation}
ds^2 = a^2[(1+2\phi)d\eta^2+2B_{,i}dx^{\mu}d\eta-((1-2\psi)\delta_{\mu\nu}-2E_{,\mu\nu})dx^{\mu}dx^{\nu}].
\label{g-inv}
\end{equation}

Under the gauge (coordinate) transformation 
\begin{equation}
x^{\mu} \rightarrow x^{\mu} + \xi^{\mu},
\label{trans}
\end{equation}
where
\begin{equation}
\xi^{\mu} = (\xi^0,\zeta^{,i})
\label{g-inv1}
\end{equation}
and $\xi^0$ and $\zeta$ are arbitrary scalar functions, 
the scalar functions in \eqref{g-inv} are transformed as follows:
\begin{equation}
\phi \rightarrow \phi-\frac{1}{a}(a\xi^0)',
\label{g-inv2}
\end{equation}
\begin{equation}
B \rightarrow B+\zeta'-\xi^0,
\label{g-inv3}
\end{equation}
\begin{equation}
\psi \rightarrow \psi+\frac{a'}{a}\xi^0,
\label{g-inv4}
\end{equation}
\begin{equation}
E \rightarrow E + \zeta.
\label{g-inv5}
\end{equation} 

Let us emphasize that 
the representation of the metric coefficients \eqref{g-inv} is general and does not depend on the choice between the gauge invariant formalism and the formalism connected with some gauge fixing. Then, we can consider the general scalar gauge transofrmations \eqref{trans}-\eqref{g-inv1}. The equations \eqref{g-inv2}-\eqref{g-inv5} show how the coefficients $B,\phi,\psi$ and $E$ are changed under the diffeomorphisms \eqref{trans}-\eqref{g-inv1}.

The functions
\begin{equation}
\Phi = \phi-\frac{1}{a}[a(B-E')]'
\label{Phi}
\end{equation}
and 
\begin{equation}
\Psi = \psi+\frac{a'}{a}(B-E')
\label{Psi}
\end{equation}
are gauge-invariant. 
The gauge-invariant expression for the perturbation of energy density is 
\begin{equation}
\overline{\delta\varepsilon}= \delta\varepsilon -\varepsilon_0'(B-E').
\label{en-inv}
\end{equation}
The gauge-invariant expressions for the components of the velocity are 
\begin{equation}
\overline{\delta u_0}=\delta u_0-[a(B-E']',
\label{vel}
\end{equation}
\begin{equation}
\overline{\delta u_{\mu}}=\delta u_{\mu}-a(B-E')_{,\mu}. 
\label{vel1}
\end{equation}
All these quantities are invariant with respect to arbitrary gauge transformations independently of the chosen gravity theory.

In General Relativity, one can choose the longitudinal (conformal-Newtonian) gauge, requiring 
\begin{equation}
B = 0,\ E =0.
\label{long}
\end{equation}
In this case 
\begin{equation}
\Phi = \phi,\ \Psi = \psi.
\label{long1}
\end{equation}
This gauge has two distinguishing features. Firstly, we can always choose two functions $\xi^0$ and $\zeta$ to eliminate 
$B$ and $E$. Indeed, if we have nonzero $B$ and $E$, we can choose 
\begin{equation}
\zeta = -E,\ \xi^0 = B-E'.
\label{long2}
\end{equation}
Secondly, when the gauge is fixed we do not have a residual freedom, permitting to conserve the conditions \eqref{long}.
Note, that when one uses the synchronous gauge $\phi = 0,\ B = 0$, there is large residual freedom (see e.g. \cite{Land-Lif}). 

Now, we would like to consider the generalized unimodular gravity with the radiation-type condition 
\begin{equation}
N = \gamma^{\frac{1}{6}}.
\label{com}
\end{equation}
What will be shown for this particular case is valid also for a more general condition $N = \gamma^{n}$, but we choose  condition \eqref{com} for two reasons:
firstly, it is interesting from the physical point of view, secondly, the corresponding formulas look simpler.

In the case of the generalized unimodular gravity, we cannot require simultaneously $B= 0$ and $E = 0$, because we already have a fixed gauge condition, which in the radiation-like case implies 
\begin{equation}
\phi = -\psi - \frac{\Delta E}{3}.
\label{uni-gauge}
\end{equation}
However, we still can require $B = 0$. To combine this condition  with the condition \eqref{uni-gauge} we should  use the functions $\xi^0$ and $\zeta$, satisfying the equations 
\begin{eqnarray}
&&(\xi^{0})'+\frac{\Delta \zeta}{3} = \phi+\psi+\frac{\Delta E}{3},\nonumber \\
&&\zeta'-\xi^0=-B.
\label{uni-gauge1}
\end{eqnarray}
Meanwhile we can use the gauge-invariant variables to write down the Einstein equations. We shall have the effective pressure and effective energy density, corresponding to the presence of the radiation in our model.   

When the universe is filled with a perfect fluid the spatial components of the energy-momentum tensor are diagonal and from this fact it follows that 
\begin{equation}
\Psi = \Phi.
\label{phi-psi}
\end{equation}
Now, we can write down the following set of the effective Einstein equations:
\begin{equation}
\Delta \Phi - 3{\cal H}(\Phi'+{\cal H}\Phi) = 4\pi Ga^2\overline{\delta\varepsilon},
\label{Ein}
\end{equation}
\begin{equation}
(a\Phi)'_{,\mu} = 4\pi G a^2(\varepsilon_0+p_0)\overline{\delta u_{\mu}},
\label{Ein1}
\end{equation}
\begin{equation}
\Phi''+3{\cal H}\Phi'+(2{\cal H}'+{\cal H}^2)\Phi = 4\pi Ga^2\overline{\delta p}.
\label{Ein2}
\end{equation}
Here 
\begin{equation}
{\cal H} = \frac{a'}{a}
\label{H}
\end{equation}
and in the case of the radiation
\begin{equation}
a \sim \eta
\end{equation}
and 
\begin{equation}
{\cal H} = \frac{1}{\eta}.
\end{equation}
Combining Eqs. \eqref{Ein} and \eqref{Ein2}, we obtain the following equation
\begin{equation}
\Phi''+4{\cal H}\Phi'-\frac13\Delta\Phi=0.
\label{Ein3}
\end{equation}
For the Fourier plane-wave harmonics
\begin{equation}
\Phi = \Phi_{\vec{k}}e^{i\vec{k}\cdot\vec{x}},
\label{Fourier}
\end{equation}
Eq. \eqref{Ein3}
looks as follows:
\begin{equation}
\Phi_{\vec{k}}''+4H\Phi_{\vec{k}}'+\frac13k^2\Phi_{\vec{k}}=0.
\label{Ein4}
\end{equation}
Its general solution is 
\begin{equation}
\Phi_{\vec{k}} = \frac{1}{x^2}\left[C_1\left(\frac{\sin x}{x}-\cos x\right)+C_2\left(\frac{\cos x}{x}+\sin x\right)\right],
\label{Ein5}
\end{equation}
where 
\begin{equation}
x \equiv \frac{k\eta}{\sqrt{3}}.
\label{x}
\end{equation}
Now, let turn our attention to Eq. \eqref{Ein1}. It is a time-space component of the effective Einstein equations. 
However, in our version of the generalized unimodular gravity, the variation of the action with respect to the shift functions 
gives zero, i.e. the mixed time-space components of the energy-momentum tensor are zero, or, the velocities are zero.
Nevertheless the right-hand side of Eq. \eqref{Ein1} does not vanish, because we have there not the velocity, but its gauge-invariant version (see Eq.   \eqref{vel1}). Thus, we have
\begin{equation}
(a\Phi)'_{,i} = 4\pi G a^3(\varepsilon_0+p_0)E'_{,i}.
\label{Ein6}
\end{equation}
 Then, we obtain
 \begin{equation}
 E' = C_0a(a\Phi)' + f(\eta),
 \label{Ein7}
 \end{equation}
 where 
 $C_0$ is a constant depending on the global quantity of the effective radiation in the universe, and $f(\eta)$ is a function which does not depend on spatial coordinates.
 Then,
 \begin{equation}
 E = C_0\int d\eta a(a\Phi)' + F(\eta).
 \label{Ein8}
 \end{equation}
 
 Now, we know the function $\Phi$ and the function $E$, expressed in terms of the function $\Phi$ and the unknown time-dependent function $F(\eta)$. We can find  the function $\phi$ from Eqs. \eqref{Phi} and \eqref{Ein7}:
 \begin{equation}
 \phi = \Phi -\frac{1}{a}[a(C_0a(a\Phi)'+f]'.
 \label{phi-new}
 \end{equation}
 Analogously, Eqs. \eqref{Psi}, \eqref{phi-psi} and \eqref{Ein7} give
 \begin{equation}
 \psi = \Phi +\frac{a'}{a}(C_0a(a\Phi)'+f).
 \label{psi-new}
 \end{equation}
 Substituting the expressions \eqref{Ein8}, \eqref{phi-new} and \eqref{psi-new} into Eq. \eqref{uni-gauge}, we obtain the following consistency condition:
 \begin{equation}
 2\Phi-C_0(a(a\Phi)')'-f'+\frac{C_0}{3}\int d\eta a(a\Delta\Phi)'=0.
 \label{consist}
 \end{equation}

It is more convenient to rewrite these equations in terms of the conformal time $\eta$, eliminating the constant $C_0$ and $a$.
We obtain 
\begin{equation}
E'_{,i} = \frac12\eta(\eta\Phi)'_{,i},
\label{eta}
\end{equation}
\begin{equation}
E'=\frac12\eta(\eta\Phi)'+f(\eta),
\label{eta1}
\end{equation}
\begin{equation}
E =\frac12 \int d\eta \eta(\eta\Phi)'+F(\eta).
\label{eta2}
\end{equation}
\begin{equation}
\phi = \Phi -\frac{1}{\eta}\left[\eta\left(\frac12\eta(\eta\Phi)'+f\right)\right]',
\label{eta3}
\end{equation}
\begin{equation}
\psi = \Phi+\frac{1}{\eta}\left(\frac12\eta(\eta\Phi)'+f\right),
\label{eta4}
\end{equation}
\begin{equation}
2\Phi-\left(\frac12\eta(\eta\Phi)'+f\right)'+\frac16\int d\eta\eta(\eta\Delta \Phi)' = 0.
\label{consist1}
\end{equation}
Substituting into Eq. \eqref{consist1} the expression for the Laplacian from Eq. \eqref{Ein3}, we obtain the expression 
which depends only on the function $\Phi$ and its time derivatives. After integration by parts we see that the consistency condition \eqref{consist1} is satisfied for all the solutions of Eq. \eqref{Ein3}. That means that all the solutions arising in the General Relativity in terms of gauge-invariant potentials are valid also for the generalized unimodular gravity, but their interpretation is different because we cannot connect the gauge-invariant potentials with the Newtonian potential since we cannot introduce the conformal-Newtonian gauge.\\

\section{Conclusion}

The usual unimodular condition, for which the determinant of the metric is fixed as a constant, can be generalized by considering the lapse function $N$ as a function of the determinant of the spatial part of the metric $\gamma$.
In the cosmological context, this implies to obtain a description for the dark sector of the universe in terms of geometric quantities. In the context of the present analysis, the effective equation of state of this {\it geometric} dark fluid is restricted to a linear relation between the effective mater and pressure. The usual expressions for the background universe can be re-obtained but in a pure geometric setting.

A perturbative analysis has also been made using this generalized unimodular theory. Employing the gauge invariant formalism, the usual results obtained with the GR theory are recovered. However, due to the constraint imposed on the lapse function, some new relations appear on the perturbed metric variables. In particular, the longitudinal newtonian gauge can not be used. This means that, even if the results are the same as in GR, their interpretation may differ. 

To clarify this interpretation issue it is necessary a more complete analysis related to the cosmological observables and an extension of the model developed here to a more realistic scenario, mainly by including the ordinary matter content (specially the baryonic and radiative components).
This is postpone for a future study.
\\

\noindent
{\bf Acknowledgements:} J.C.F. thanks CNPq (Brazil) and FAPES (Brazil) for partial financial support. A.Yu.K. is grateful to the 
N\'ucleo Cosmo-ufes \& Departamento de F\'isica, Universidade Federal do Esp\'irito Santo 
for the financial support and kind hospitality during his visit to Vitoria in January of 2026.

\end{document}